\documentclass[aps,prl,twocolumn,floats,nofootinbib,longbibliography,reprint,amsmath,amssymb,showpacs,amsart]{revtex4-1}
\usepackage{graphicx}
\usepackage{dcolumn}
\usepackage{bm}
\usepackage{multirow}
\usepackage{moreverb}
\usepackage{natbib}
\usepackage{textcomp}

\begin{document}

\preprint{APS/123-QED}

\title{Generation of very high-order high purity Gaussian modes via spatial light modulation}

\author{Stefan Ast}
\email{stefan.ast@aei.mpg.de}
\thanks{Present Address: Max Planck Institute for Gravitational Physics (Albert Einstein Institute) and Leibniz Universit\"at Hannover, Germany}
\author{Sibilla Di Pace}%
\thanks{Present Address: Università degli Studi di Roma La Sapienza \& INFN Sezione di Roma1, Rome, Italy}
\author{Jacques Millo}
\thanks{Present Address: Institut FEMTO-ST, Besançon, France}
\author{Mikhael Pichot}
\author{Margherita Turconi}
\author{Walid Chaibi}
\email{chaibi@oca.eu}
\affiliation{ARTEMIS, Universit\'e C\^ote d\textquotesingle Azur, CNRS and Observatoire de la C\^ote d\textquotesingle Azur, Boulevard de l'Observatoire F-06304 Nice, France}

\date{\today}

\begin{abstract}
We experimentally demonstrate the conversion of a fundamental $\text{TEM}_{00}$  laser mode at 1064\,nm to higher order Hermite-Gaussian modes (HG) of arbitrary order via a commercially available liquid crystal Spatial Light Modulator (SLM). We particularly studied the  $\text{HG}_{5,5}/\text{HG}_{10,10}/\text{HG}_{15,15}$ modes. A two-mirror plano-spherical cavity filters the higher-order modes spatially. We analyze the cleaned modes via a three-mirror diagnosis cavity and measure a mode purity of 96/93/78\% and a conversion efficiency of 6.6\%/3.7\%/1.7\% respectively. The generated high-purity Hermite-Gaussian modes can be employed for the mitigation of mirror thermal noise in optical cavities for both optical clocks and gravitational wave (GW) detectors. HG modes are then converted into high order LG modes which can be of particular interest in cold atom physics.

\end{abstract}

\pacs{Valid PACS appear here}
\maketitle

Lasers used for high-precision experiments typically generate an output beam in the fundamental $\text{TEM}_{00}$ mode, since an operation at higher-order $\text{HG}_{m,n}$ or $\text{LG}_{p,l}$ modes suffers from low lasing efficiency due to diffraction losses and a less stable output mode~\cite{siegman1986lasers}. These are, however, benificial in metrology experiments using optical cavities or laser interferometers which are fundamentally limited by thermally induced mirror surface motions that reduce the length sensing sensitivity \cite{Thermal-noise-theory-2004}. More specifically, reference cavities used for the frequency stabilization of the optical oscillator in optical clocks are currently limited in stability by thermal noise of the highly-reflective mirror coatings with a fractional frequency instability of $4\times 10^{-17}$ in 1\,s~\cite{Matei2017}. Advanced GW detectors are also limited by the coatings' thermal noise at $\text{h}=10^{-23}[1/\sqrt{\text{Hz}}]$ in the frequency band of 30-300\,Hz \cite{GWD-Thermal-noise-2002,ALIGO-2015}, which will limit their sensitivity and the sensitivity of the third generation GW detectors \cite{Einstein-telescope-2010}.
The negative effect of thermal noise can be mitigated by using a spatially broader intensity profile for the laser beam in comparison to the fundamental $\text{TEM}_{00}$ mode. This results in a higher averaging over the mirror surface for the same cavity characteristics. Using flat-top beams was initially proposed~\cite{Ambrosio2003,Ambrosio2004_2} and its efficiency to reduce thermal and thermoelastic noises was theoretically demonstrated~\cite{O'Shaughnessy2004,Ambrosio2004,Vinet2005}. However, the experimental implementation of these modes is troublesome since it implies the use of the so-called Mexican-hat mirrors. Mitigation of thermal noise can also be achieved by higher-order modes whose efficiency increases with the total mode order $N$ ($N = m+n$ for HG$_{m,n}$ modes and $N = 2l+p$ for LG$_{p,l}$ modes)~\cite{Vinet2010}. Higher-order Laguerre-Gauss modes were the first to be suggested for ground based GW detectors~\cite{ALIGO-2015,Mours2006} and this idea was extended for reference cavities related to optical clocks \cite{best-optical-clock-2013}. For GW detection, the high order modes must be: high-power (efficiently generated), pure and robust with respect to mirrors defects and aberrations. While the LG$_{3,3}$ mode has been the object of several experimental investigations in this context~\cite{Granata2010,Gatto2014,Fulda2010,Bond2011, Allocca2015,carbone2013,Chelkowski2009,Hong2011,Sorazu2013,Noack2017}, there has not yet been any experiment concerning high order HG modes. A reduction of thermal noise was recently demonstrated in a 10-cm Ultra Low Expansion (ULE) glass reference cavity using the HG$_{0,2}$ mode. Furthermore, LG$_{0,l}$ modes are of particular interest for capturing and guiding cold atoms which have a significant impact on quantum optics, quantum gases physics and metrology~\cite{Grier2003,Grier2006,Nogrette2014}. They provide a donut intensity shape, which remains unchanged while propagating.  This shape allows transverse trapping of cold atoms in a potential well which is even more steep with increasing $l$~\cite{Viaris2007,Mestre2010,Carrat2014}. LG$_{0,l}$ modes can also be used to transfer multi $\hbar$ angular momentum~\cite{Allen1992}, i.e. link atomic levels with different $m_{F}$. For those applications, the mode purity is critically important. In this letter we investigate the conversion of the fundamental $\text{TEM}_{00}$ mode at 1064\,nm to higher-order high purity $\text{HG}_{m,n}$ and $\text{LG}_{p,l}$ modes for the very first time.

\subparagraph{LG versus HG}
The thermal noise at low frequencies in the case of an optical cavity operating with axisymmetric LG modes is computed in~\cite{Mours2006,Vinet2009} using the Bondu-Hello-Vinet (BHV) technique~\cite{Bondu1998}(amended in~\cite{Liu2000}). This calculation is based on Levin's theory~\cite{Levin1998} and is applied on finite mirrors. HG modes' related thermal noise was computed analytically in~\cite{Vinet2010} only in the case of infinite mirrors. Based on the latter reference, we compare in Tab.\ref{table_mode_efficiency}, the mitigation efficiency of thermal noise for the LG$_{5,5}$, a frequent example in theoretical computations~\cite{Mours2006,Vinet2007,Vinet2009}, the LG$_{3,3}$ and the HG modes presented in this letter: HG$_{5,5}$, HG$_{10,10}$. Coating Brownian thermal noise is currently limiting both reference cavities and GW detectors, whereas the thermoelastic noise limit, which undergoes similar reduction factors, lies below that and is not considered here. LG modes are more efficient to mitigate thermal noise for equivalent clipping losses. However, it was found that the LG modes pose strong requirements on the mirror surface quality since they exhibit a mode pseudo-degeneracy~\cite{Hong2011} which greatly decreases their coupling efficiency to linear cavities. Residual astigmatism of the mirrors turns out to be the main defect responsible for the intracavity mode degradation~\cite{Bond2011,Sorazu2013} and its thermal compensation was demonstrated to improve the mode quality~\cite{Allocca2015}. Astigmatism also prevents LG modes from resonating in a 4 mirror pre-mode cleaner unless in a non-planar configuration~\cite{Noack2017}. Note that LG modes
do not resonate in triangular cavities~\cite{Fulda2010} which are largely used in GW detectors.

\begin{table}[h]
   \caption{Thermal noise limit reduction for higher order modes with respect to the Gaussian mode for both substrate and coatings. $a$ represents the mirror radius and $w$ designates the beam radius. The ratio $a/w$ considered here corresponds to a clipping loss of 1 ppm.}
	\label{table_mode_efficiency}
	{\renewcommand{\arraystretch}{1} 
{\setlength{\tabcolsep}{0.4cm} 
\begin{tabular}{c c c c c}
  \hline\hline
  & order & Substrate & Coatings & $a/w$ \\
  \hline
  \multirow{2}{*}{HG} & 5,5 & 0.576 & 0.493 & 4.379 \\
    & 10,10  & 0.491 & 0.392 & 5.513 \\ \hline
  \multirow{2}{*}{LG} & 3,3  & 0.559 & 0.378 & 3.313 \\
    & 5,5  & 0.496 & 0.310 & 4.968 \\ 
  \hline\hline
	
\end{tabular}}}
\end{table}
Whereas astigmatic LG modes are not a mathematical solution of the free propagation equation, astigmatic HG modes are, and they are likely to couple into a 2-mirror astigmatic cavity (the case of 3-mirror cavities will be described below). To that end, one needs to angle the two mirrors around the optical axis in order to match their own axes to those of the injected HG mode. To confirm these assumptions, FFT based simulations using the software DarkF~\cite{Virgo2007} on a 3km long cavity of a finesse 1000 involving an Advanced Virgo like mirror (surface roughness 0.5 nm rms, radius of curvature 3465 m) and a perfect flat mirror were conducted. LG$_{3,3}$ and LG$_{5,5}$ mode coupling efficiency does not exceed 70\%. On the other hand, the HG$_{5,5}$ and HG$_{10,10}$ mode coupling increases from 77\% to 99\% and from 62\% to 96\% respectively when one rotates the mirrors by the right angle(see supplementary material).This renders higher-order HG modes as promising candidates for applications in GW detectors and reference cavities for optical clocks, although they grant a slightly smaller mitigation effect of thermal noise.
\begin{figure}[h]
\centering
\resizebox{8.5cm}{!}{\rotatebox{0}{\includegraphics[width=\linewidth]{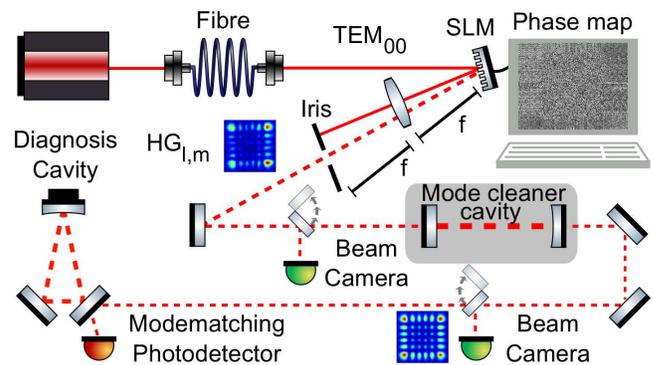}}}
\caption{Simplified scheme of the experimental setup. The fundamental laser mode at 1064\,nm is spatially filtered via a single-mode polarization maintaining fiber. A computer generated phase map of the higher order $\text{HG}_{l,m}$ mode is applied to the liquid crystal spatial light modulator (top right). An iris seperates the higher-order mode from the $\text{TEM}_{00}$ mode via a Fourier lens with a focal length f. The higher order laser beams ($\text{HG}_{5,5}/\text{HG}_{10,10}/\text{HG}_{15,15}$) are spatially filtered via a two mirror cavity. The modeshape is detected with a beam camera and the mode-purity is analyzed with a three-mirror cavity (Diagnosis Cavity). A custom-made photodetector measures the modematching of the diagnosis cavity in reflection.}
\label{fig:exp-setup-pic}
\end{figure}

\subparagraph{Generation of HG modes}
In our experiment we achieved the conversion of a fundamental $\text{TEM}_{00}$ mode to higher-order Hermite-Gaussian modes via the reflection on a commercially available computer addressed liquid crystal phase-only SLM by \emph{Hamamatsu Photonics}. The same technique was used by several groups to generate LG beams for different purposes \cite{LG-modes-via-liquid-crystal-SLM-2008,LCOS-2007,Ando2009,Ohtake2007,Rhodes2006,Fatemi2006,Mestre2010,Carrat2014}. The amplitude distribution of the reflected beam is then conserved whereas the phase distribution is exclusively defined by the phase map $\varphi_{\textrm{SLM}}\left(x,y\right)$. The higher-order mode is then obtained out of the field distribution $\textrm{E}_{\text{out}}$ in the image focal plane of a Fourier lens. This field distribution is thus related to the incident beam according to $\mathcal{F}\left[ \text{E}_{\text{out}} \left(X,Y\right)\right] = \lvert\text{TEM}_{00}\left(x,y\right) \rvert \exp\left[ i \varphi_{\text{SLM}}\left(x,y\right)\right]$ where $\mathcal{F}$ denotes the Fourier transform. Since a $\text{HG}_{m,n}$ mode is invariant by $\mathcal{F}$, the last equation shows that a fundamental Gaussian mode cannot be converted into an arbitrary $\text{HG}_{m,n}$ mode with 100\% efficiency. In \cite{Pavelyev-1998,Duparre-1998} it is proposed to generate the $\text{HG}_{m,n}$ mode in a limited area of the Fourier plane while no condition is imposed on the rest of the plane. The approach is inspired from the Gerchberg-Saxton algorithm \cite{Gerchberg1972}. It consists of consecutive iterations between the SLM and the Fourier planes using respectively the Fourier and inverse Fourier transforms while imposing on each iteration the fundamental mode amplitude on the SLM and the $\text{HG}_{m,n}$ mode in the restricted area of the Fourier plane. Originally, the Gerchberg-Saxton algorithm imposes an amplitude condition in the Fourier plane and was successfully used for different purposes (see for example \cite{Nogrette2014} and the references therein). Different conditions can be imposed provided enough degrees of freedom are respected \cite{Pasienski2008}. In our case, both amplitude and phase are imposed in the Fourier plane but in a restricted area. This iterative approach was generalized by Levi et. al. and received the name of generalized projection method \cite{Generalized-projections-Levi-1983}. In iterative Fourier transform algorithms, vortex stagnation \cite{Senthilkumaran2005} issues might occur which correspond to a very slow convergence or its absence. Relaxation parameters can be introduced to the algorithm to speed up its convergence \cite{Generalized-projections-Levi-1983} and was used for the generation of a $\text{HG}_{1,0}$ mode in refs.\cite{Pavelyev-1998,Duparre-1998}. No such effect was observed in our case and relaxation parameters were tested to be unnecessary. 20 iterations were sufficient to obtain an arbitrary order $\text{HG}_{m,n}$ mode within less than 1\% error and with a theoretical conversion efficiency spreading from 45\% for the $\text{HG}_{5,5}$ mode to 20\% for the $\text{HG}_{15,15}$ mode. The SLM diffraction efficiency decreases with the spatial frequency of the phase map, a non-diffracted light was superimposed to the HG mode in the Fourier plane. A blazed grating was then added to the SLM phase map in order to obtain a free diffraction area. Unwanted light (undiffracted and diffracted in the free area) was spatially filtered with an iris placed in the Fourier plane.\\ 

A simplified schematic picture of the experimental setup is given in Fig.~\ref{fig:exp-setup-pic}. The laser source is a continuous-wave single frequency Nd:YAG non-planar ring oscillator (NPRO) at 1064\,nm. The laser beam is spatially filtered via a single-mode polarization maintaining fiber. A fiber-based electro-optical modulator generates phase modulated sidebands at 100\,MHz for a Pound-Drever-Hall (PDH)~\cite{Drever1983} cavity stabilization scheme. The phase map generated by the iterative algorithm is then applied to a liquid crystal phase-only SLM type LCOS-SLM X10468 having $792\times600$ pixels with a pixel pitch of $20\,\mu \text{m}$. We used a laser power of about 3\,mW and up to 1 W on the SLM. An iris in the image focal plane of the Fourier lens serves as a spatial frequency filter for the residual light from the conversion process. The $\text{HG}_{5,5}/\text{HG}_{10,10}/\text{HG}_{15,15}$ modes are transmitted respectively to a 15 cm long two mirror plano-spherical cavity acting as a mode cleaner. It spatially cleans the higher-order mode from residual light in unwanted transversal modes by stabilizing the laser frequency to the cavity's length via the PDH scheme. The mode cleaner cavity Finesse is 457 with a free spectral range of 997\,MHz. We achieved a mode-matching of 62\% for the $\text{HG}_{5,5}$ mode, 47\% for the $\text{HG}_{10,10}$ mode and 28\% for the $\text{HG}_{15,15}$ mode, respectively.\\
For high spatial frequency diffraction as present in our experiment, the SLM's expected efficiency from its data-sheet  is limited to a maximal value of 40\%. The iterative algorithm calculated a projection efficiency from the $\text{TEM}_{00}$ mode to the $\text{HG}_{5,5}$  of about 28\% and to the $\text{HG}_{10,10}/\text{HG}_{15,15}$ modes of about 25\%. These values are used to achieve a trade-off between a technically feasible mode size and the SLM's pixel size. Taking into account the mode coupling into the mode cleaner cavity, the total conversion efficiencies are expected to be 6.9\% for the $\text{HG}_{5,5}$ mode, 4.7\% for the $\text{HG}_{10,10}$ mode and 2.8\% for the $\text{HG}_{15,15}$. The experiment achieved a conversion efficiency of respectively 6.6\%/3.7\% and 1.7\% for $\text{HG}_{5,5}/\text{HG}_{10,10}/\text{HG}_{15,15}$ which slightly decreases with an increase of the total mode order $N=m+n$. For these modes, the gap between measured conversion efficiencies and corresponding expected values is satisfactory, although it increases with the total mode order for unknown reasons.\\

\begin{figure}[ht]
\centering
\resizebox{8cm}{!}{\rotatebox{0}{\includegraphics[trim=25mm 85mm 20mm 85mm, clip=true]{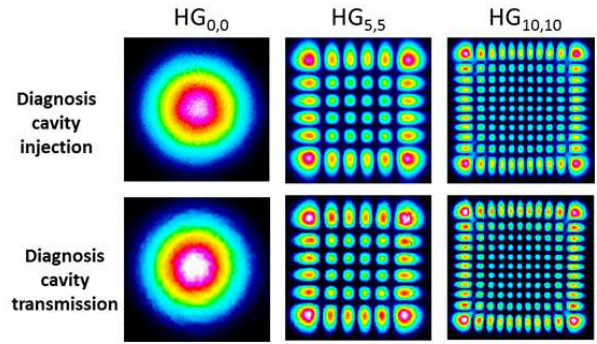}}}
\caption{Comparison between the injected and transmitted beams through the triangular diagnosiss cavities for HG$_{0,0}$, HG$_{5,5}$ and HG$_{10,10}$ modes showing that the cavity mode is a true HG mode. Transmitted images are slightly fuzzy because they were taken while the diagnosis cavity is being scanned.}
\label{fig:injection_transmission}
\end{figure}

\subparagraph{Beam purity measurement}
\begin{figure}[h]
\centering
\resizebox{8cm}{!}{\rotatebox{0}{\includegraphics[trim=15mm 15mm 15mm 15mm, clip=true]{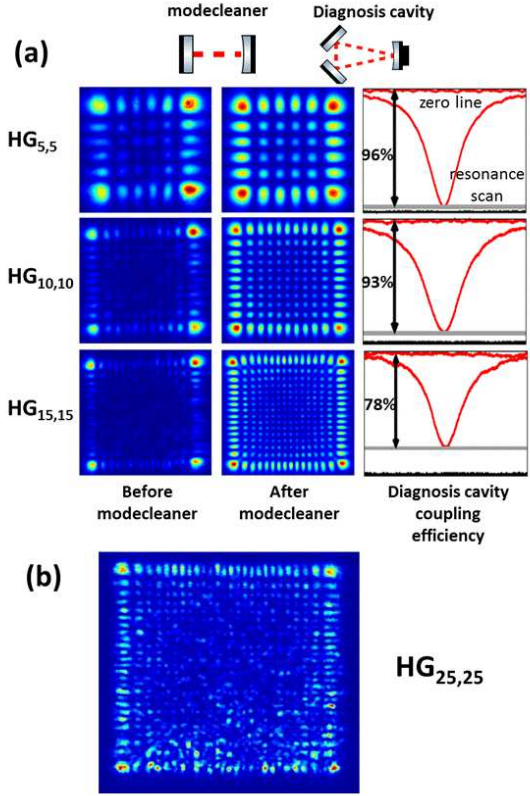}}}
\caption{(a) Intensity profile measurements of the $\text{HG}_{5,5}$, $\text{HG}_{10,10}$ and $\text{HG}_{15,15}$ modes (bottom) using a \emph{Dataray WincamD} beam camera. The mode purity is given by the measurement of the coupling efficiency on the the reflection signal of  the three-mirror diagnosis cavity. The red resonance curves depict a low amplitude length scan over the diagnosis cavity resonance. (b) Beam camera measurement showing a $\text{HG}_{25,25}$ mode. The picture was taken directly after the generation of the mode without spatial filtering, since the mode's low purity excluded a sufficient mode-matching and mode cleaner cavity stabilization.}
\label{coupling_HG2525}
\end{figure}

As pointed out by Fulda \textit{et. al.}~\cite{Fulda2010}, one can inject the mode into a non-degenerate diagnosis cavity and measure its coupling efficiency in order to quantify its purity. This method was identified to be non ideal to test the purity of LG modes because the cavity modes themselves are not pure LG modes. A partial measurement of the LG mode purity was made by fitting the intensity profile with the corresponding analytical function at some spatial location~\cite{Fulda2010,Granata2010}. Unless the mirrors are angled with respect to each other and to the incoming mode as described above, testing HG modes faces the same difficulty. However, at some level of the cavity astigmatism, the degeneration lifting between modes of constant total order $N=m+n$ exceeds the cavity bandwidth, which then cleans up the cavity mode. FFT based simulations were conducted on the same Virgo-like cavity by artificially adding a curvature to the end mirror either along $x$ or the $y$ axis. Coupling efficiencies exceeding 97\% for both HG$_{5,5}$ and HG$_{10,10}$ are obtained (see supplementary material).\\      
We analyzed the mode-purity via a three-mirror triangular cavity whose mode is naturally astigmatic. It consists of two plane mirrors and a high-reflective spherical back mirror having a cavity Finesse about 709. The cavity back-mirror is attached to a piezo-electric actuator to tune the cavity length. The roundtrip length of the cavity is 42.5\,cm, resulting in a free spectral range of 706\,MHz. The beam impinges on the 50 cm radius of curvature spherical mirror with an angle of $4^\circ$. This makes the degeneracy lifting between modes of constant total order approximately 1.75 times the cavity pole. The triangular cavity can then be considered as a HG mode reference. Note that axisymmetric LG modes do not resonate in triangular cavities~\cite{Fulda2010}. The mode was detected in transmission of the diagnosis cavity via a \emph{Dataray WincamD} beam camera ($512\times 512$ pixels and a pixel size of $9.3\times 9.3\,\mu \textrm{m}^{2}$) while scanning the cavity length. In Fig.~\ref{fig:injection_transmission} the cavity transmitted beam is displayed for different injected modes and appear to be qualitatively pure HG modes.\\

The mode-matching of the fundamental mode is measured by observing the resonance peak on the reflected signal while the cavity length is being scanned. Special care is taken to collect the whole reflected power on the photodiode. We first matched the fundamental HG$_{0,0}$ to have a reference for the achievable modematching of the higher order modes and the coupling was better than 99\%. The modematching of the higher-orders required some slight additional adjustments and revealed a coupling efficiency of 96/93/78\% for the $\text{HG}_{5,5}/\text{HG}_{10,10}/\text{HG}_{15,15}$ modes as depicted in Figure~\ref{coupling_HG2525}.a. In addition, we generated a $\text{HG}_{25,25}$ mode to show the possibility of generating arbitrarily high-order modes. Its intensity profile right after the Fourier plane is shown in Fig.~\ref{coupling_HG2525}.b. We were not able to achieve a sufficient coupling to stabilize the $\text{HG}_{25,25}$ mode to the modecleaner cavity due to its low mode purity related to the SLM finite spatial frequency and the extremely high order mode sensitivity to mismatching and misalignment. However, these are technical problems and will not pose any fundamental limitation in the generation of modes as high as the $\text{HG}_{25,25}$ mode or even higher modes.\\

\subparagraph{HG to LG conversion}

\begin{figure}[h]
\centering
\resizebox{7.5cm}{!}{\rotatebox{0}{\includegraphics[trim=15mm 40mm 15mm 40mm, clip=true]{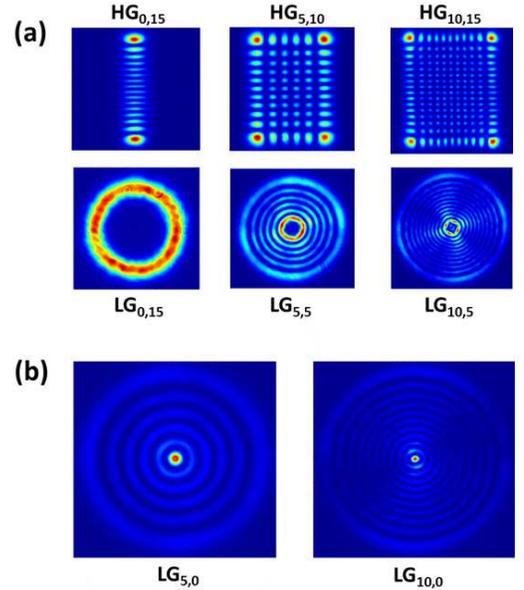}}}
\caption{(a) LG modes generated from corresponding HG modes. The HG mode shown here are imaged after non tilted cylindrical lenses.(b) LG modes generated from pure HG$_{5,5}$ and HG$_{10,10}$ modes. The central peak is typical of LG$_{p,0}$ modes.}
\label{HG_LG_conversion}
\end{figure} 

The high purity HG modes we generate here are converted into LG modes using a simple combination of two cylindrical lenses ~\cite{Tamm1990,Beijersbergen1993,Padgett1996}. HG modes form a complete sets of solutions on which any propagation mode can be decomposed. Interestingly, $45^\circ$ tilted HG$_{m,n}$ mode have the same decomposition on the straight HG modes set as the LG$_{\textrm{min}\left(m,n\right),m-n}$ modes except for a multiple of $\pi/2$ phase for each term~\cite{Abramochkin1991}. When the $45^\circ$ tilted mode crosses two identical cylindrical lenses placed around its waist, with the conditions $f = \sqrt{2}\,d$ and $z_R = f+d$ ($f$ is the focal length of the lenses, $d$ is the distance separating them and $z_R$ is the Rayleigh range of the impinging mode), each term of the decomposition on the straight modes set accumulates a Gouy phase shift corresponding to the required multiple of $\pi/2$ allowing the conversion of the $45^\circ$ tilted HG mode to the corresponding LG mode~\cite{Beijersbergen1993}. In Fig.\ref{HG_LG_conversion}.a, we show a LG$_{0,15}$ generated from a HG$_{0,15}$ mode which, to our knowledge, is the highest donut mode order ever generated offering an important angular momentum transfer. LG$_{5,5}$ and LG$_{10,10}$ are also generated and we are currently studying their behavior in a middle finesse cavity and comparing them with HG modes. This technique can also be used the other way around: LG modes can be generated by the SLM and converted into HG mode before being injected into the mode cleaner. This profits from high efficiency generation of LG modes (tens of \%~\cite{Mestre2010}) to overcome the lack of efficiency in direct generation of HG modes. In Fig.\ref{HG_LG_conversion}.b, both LG$_{5,0}$ and LG$_{10,0}$ are generated from pure HG$_{5,5}$ and HG$_{10,10}$ to show the match between these particular modes. Except for this, no clear application is known for LG$_{p,0}$ modes since they do not carry any angular momentum and they have high intensity central peaks preventing the thermal noise mitigation.

\subparagraph{Conclusion}
We experimentally demonstrated the transformation of the fundamental $\text{TEM}_{00}$ mode to higher-order HG laser modes as high as $\text{HG}_{25,25}$ via a liquid crystal spatial-light modulator. The generated $\text{HG}_{5,5}/\text{HG}_{10,10}/\text{HG}_{15,15}$ modes were filtered via a two-mirror modecleaner cavity and analyzed by a three-mirror triangular cavity. The experiment achieved a conversions efficiency of 6.6/3.7/1.7\%  and a mode purity of 96/93/78\% for the $\text{HG}_{5,5}/\text{HG}_{10,10}/\text{HG}_{15,15}$ modes. High order HG modes were converted into axisymmetric LG modes using a system of two $45^\circ$ tilted cylindrical mirrors. To our knowledge, this is the first time modes of such high order mode and purity were published. \\
Following these results, we conclude that the generated $\text{HG}_{l,m}$ modes are compatible with the frequently used  triangular three-mirror cavities used in GW detectors. Whereas three mirror cavities accommodate pure, rather slightly astigmatic, HG eigenmodes, linear two-mirror cavities suffer from mode pseudo degeneracy related to mirrors aberration. In comparison to LG modes, HG modes exhibit the advantage of adapting themselves to the cavity astigmatism provided the cavity mirrors are properly angled to each other and to the impinging mode. HG modes are then less sensitive to mirrors aberrations which is highly beneficial when implemented in metrology experiments~\cite{Sorazu2013,Noack2017}. Our results thus pave the way for further investigating higher-order HG modes for the reduction of thermal noise in optical cavities. 

\begin{acknowledgments}
We gratefully thank Nelson Christensen for reviewing this letter and Laurence Pruvost for interesting discussions. This work was funded by Agence Nationale de la Recherche (ANR) (ANR-09-BLAN-0207) and Labex First-TF.
\end{acknowledgments}
    
\bibliography{HGbib}
\clearpage
\appendix
\section{Supplementary Material : FFT based simulations of high order propagation modes behavior in a Virgo like real cavity}

\begin{figure}[ht]
\centering
\resizebox{8cm}{!}{\rotatebox{0}{\includegraphics[trim=15mm 20mm 15mm 20mm, clip=true]{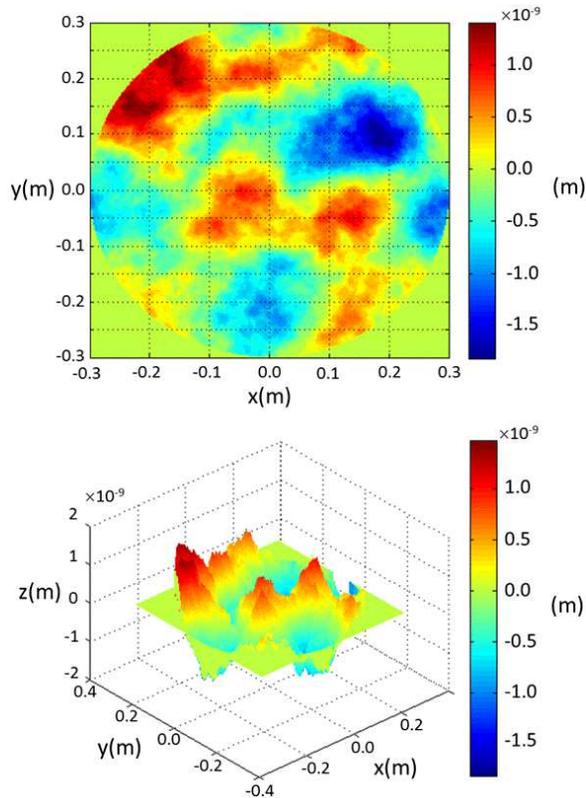}}}
\caption{Typical coated mirror distortion when piston, tilts and curvature are subtracted in the AdvancedVirgo project. The residual error here is 0.5 nm rms. }
\label{mirror_map}
\end{figure}

We present here the FFT based simulations of a 3 km plano-concave cavity with a finesse 1000. The input mirror is plane and has a transmission $T=3137\,\textrm{ppm}$. Its diameter is 30 cm. The end mirror has an identical transmission and has a radius of curvature $R=3464\,\textrm{m}$. Hence, the $g$ factor of the cavity is $g\simeq 0.11$, the cavity mode has its waist on the input mirror with a beam radius $w_0\simeq1.88\,\textrm{cm}$ and the beam radius on the end  mirror is $w_1\simeq 5.72\,\textrm{cm}$ for a laser wavelength $\lambda=1064\,\textrm{nm}$. The theoretical power enhancement factor (PEF) is $F/\pi\simeq 318$. The simulations presented below were conducted considering the input mirror as perfect whereas the end mirror presents the distortions displayed on Fig.\ref{mirror_map} which represent a Virgo like mirror surface residual when piston, tilts and curvature are subtracted. The surface error considered here is 0.5 nm rms corresponding to the requirements of the Advanced Virgo project.\\
\begin{figure}[h]
\centering
\resizebox{6cm}{!}{\rotatebox{0}{\includegraphics[trim=0mm 45mm 0mm 45mm, clip=true]{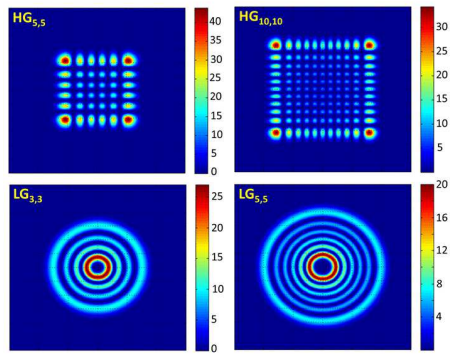}}}
\caption{Intensity distribution of the high order modes which are injected into the cavity.}
\label{perfect_modes}
\end{figure}

This cavity is simulated using the DarkF software which uses an FFT algorithm to compute the Fresnel propagation of the laser beam from one mirror to the other. The injected beams used here are LG$_{3,3}$, LG$_{5,5}$, HG$_{5,5}$ and HG$_{10,10}$ whose intensity distributions are represented in Fig.\ref{perfect_modes}.\\

\subparagraph{End mirror orientation}
\begin{table}[h]
   \caption{Coupling efficiency of considered HG and LG modes for different end mirrors angles.}
	\label{amelioration_matching}
	{\renewcommand{\arraystretch}{1}
{\setlength{\tabcolsep}{0.2cm} 
\begin{tabular}{c c c c c}
  \hline\hline
  & angle ($^\circ$) & PEF & Matching (\%) & RTL (ppm) \\
  \hline
  LG$_{3,3}$ & & 218 & 68 & 0.27 \\
	LG$_{5,5}$ & & 207 & 65 & 0.3 \\
	\hline
	\multirow{2}{*}{HG$_{5,5}$} & 0  & 246 & 77 & 0.3 \\
    & 29  & 315 & 99 & 0.56 \\
  \hline
	\multirow{2}{*}{HG$_{10,10}$} & 0  & 197 & 62 & 1.4 \\
    & 50  & 305 & 96 & 3 \\
  \hline\hline
	
\end{tabular}}}
\end{table}

The end mirror is angled around the optical axis in order to align the astigmatism axis along with the HG modes. The angle is optimized so to maximize the mode matching into the cavity. Obviously, this procedure has no effect on the LG modes coupling  efficiency. The PEF, the mode matching and the Round Trip Losses (RTL) are given in Tab.\ref{amelioration_matching} for the optimum angles, and show a clear increase in HG modes coupling efficiencies above 95\% whereas LG modes matching does not exceed 70\%. PEF also increases accordingly. Note that the optimum angles for HG$_{5,5}$ and HG$_{10,10}$ are different since the latter explores a wider surface of the mirror than the former which shows that aberration determination is mode dependent. 
\begin{figure}[h]
\centering
\resizebox{7cm}{!}{\rotatebox{0}{\includegraphics[trim=0mm 0mm 0mm 0mm, clip=true]{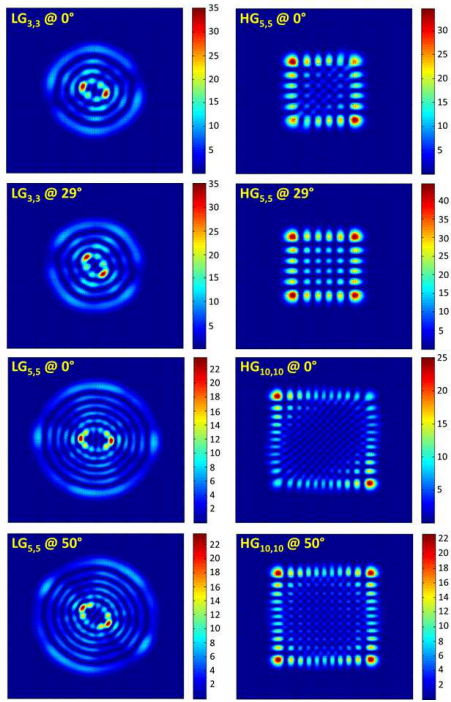}}}
\caption{Simulated intensity profiles of intracavity modes for different angles.}
\label{mirror_rotation}
\end{figure}
The simulated intensity profiles are intracavity modes are displayed in Fig.\ref{mirror_rotation}. A clear improvement of the HG modes profiles appears between $0^\circ$ angle and optimum angles for which distortions are no more observed.
\begin{figure}[h]
\centering
\resizebox{7cm}{!}{\rotatebox{0}{\includegraphics[trim=0mm 55mm 0mm 60mm, clip=true]{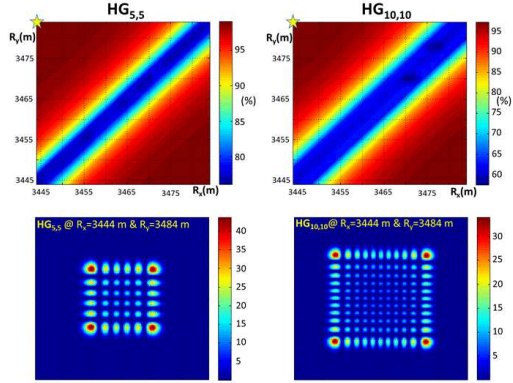}}}
\caption{\textbf{Top} : Matching as a function of mirror curvature $R_x$ and $R_y$. Yellow stars correspond to the astigmatism maximum. \textbf{Bottom} : Intensity profile of the intracavity mode for the astigmatism maximum.}
\label{mode_astigmatic_cavity}
\end{figure}
\subparagraph{Astigmatic cavity}
To mimic the astigmatism of a triangular cavity, we compute the mode behavior while changing the radius of curvature independently along the HG mode axis, namely $R_x$ and $R_y$ in the range $\left[3444\,\textrm{m}\,,\,3484\,\textrm{m}\right]$. Results of the mode matching are displayed on the top of Fig.\ref{mode_astigmatic_cavity}. On the diagonal, where only residual astigmatism exists which is not aligned with the $x$ and $y$ axis, the matching remains below 70\% for both HG$_{5,5}$ and HG$_{10,10}$. When the additional astigmatism is important enough, the matching exceed 95\%. On the bottom of Fig.\ref{mode_astigmatic_cavity}, the intensity profiles of intracavity modes are displayed for the maximum astigmatism considered here $R_x/R_y\simeq 0.99$.  

\end{document}